\documentclass[a4paper,fleqn,usenatbib]{mnras}

\usepackage[T1]{fontenc}
\usepackage{ae,aecompl}
\usepackage{graphicx}
\usepackage{amsmath}
\usepackage{amssymb}
\usepackage{supertabular}
\usepackage{hyperref}
\usepackage{xcolor}
\usepackage{academicons}
\usepackage[normalem]{ulem}
\usepackage{multirow}
\newcommand{\Msun}{\,M_{\odot}}
\def\fm3{\;\text{fm}^{-3}}

\title[Bayesian QS EOS analysis using the NICER data]
{Bayesian inference of quark star equation of state using the {\it NICER} PSR J0030+0451 data}

\author[Li et al.]{
A. Li$^{1}$\href{https://orcid.org/0000-0001-9849-3656}{\includegraphics[scale=0.07]{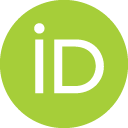}};
Z.-Q. Miao$^{1}$\href{https://orcid.org/0000-0003-1197-3329}{\includegraphics[scale=0.07]{ORCIDiD_icon128x128.png}};
J.-L. Jiang$^{2, 3}$~\href{https://orcid.org/0000-0002-9078-7825}{\includegraphics[scale=0.07]{ORCIDiD_icon128x128.png}};
S.-P. Tang$^{2,3}$~\href{https://orcid.org/0000-0001-9120-77336}{\includegraphics[scale=0.07]{ORCIDiD_icon128x128.png}};
R.-X. Xu$^{4,5}$~\href{https://orcid.org/0000-0002-9042-3044}{\includegraphics[scale=0.07]{ORCIDiD_icon128x128.png}}\\
$^1$Department of Astronomy, Xiamen University, Xiamen, Fujian 361005, China; {\tt liang@xmu.edu.cn}\\
$^2$Key Laboratory of Dark Matter and Space Astronomy, Purple Mountain Observatory, Chinese Academy of Sciences, Nanjing 210023, China\\
$^3$School of Astronomy and Space Science, University of Science and Technology of China, Hefei, Anhui 230026, China\\
$^4$School of Physics, Peking University, Beijing 100871, China\\
$^5$Kavli Institute for Astronomy and Astrophysics, Peking University, Beijing 100871, China\
}

\date{Accepted XXX. Received YYY; in original form ZZZ}

\begin{document}
\label{firstpage}
\pagerange{\pageref{firstpage}--\pageref{lastpage}}
\maketitle

\begin{abstract} 
We constrain the equation of state of quark stars within the Bayesian statistical approach using the mass and radius measurements of PSR J0030+0451 from {\it NICER}.
Three types of bag models, with and without non-zero finite quark mass and/or superfluidity, are employed for quark stars made up with self-bound strange quark matter.
We find the $90\%$ posterior credible boundary around the most probable values of the quark star maximum mass is $M_{\rm TOV}=2.38_{-0.23}^{+0.26}\Msun$, within the model flexibility of the finite quark mass, the quark pairing gap, and the perturbative contribution from the one-gluon exchange. The radius of a canonical $1.4 \Msun$ quark star is $R_{\rm 1.4}\sim12.3\,{\rm km}$, smaller than the results based on neutron star models.
\end{abstract}


\section{Introduction}           
\label{sect:intro}

The simultaneous mass ($M$) and radius ($R$) measurements of pulsar-like objects provide direct information for the phase state of the matter constituting the stars, namely the equation of state (EOS).
Supposing that the compact stars obey General Relativity, the EOS builds a unique sequence of stars in their hydrostatic equilibrium through the Tolman-Oppenheimer-Volkoff (TOV) equation.
Therefore the accurate $M$ and $R$ measurements can be used to map the EOS parameters, 
which currently cannot be calculated from the first principle theory, i.e., the quantum chromodynamics (QCD) in its non-perturbative realm~\citep[e.g.,][]{2010PhRvD..81j5021K,2014ApJ...789..127K,2018PhRvL.121t2701G}.

Promisingly, an increasing number of extremely precise mass measurements have been available, mostly in binary systems with one or two components being pulsars, by taking advantage of the extreme regularity of pulses, such as the presently heaviest PSR J0740+6620~\citep{2020NatAs...4...72C}.
Radius measurement is much more difficult~\citep{2016ARA&A..54..401O,2016RvMP...88b1001W} and possible from the pulsed emission caused by hot spots of (millisecond) pulsars, 
but previous data with large error bars do not help distinguish different EOSs until the recent
observations of {\it NICER}~\citep{2019ApJ...887L..24M,2019ApJ...887L..21R}.
After the release of the PSR J0030+0451 data from NASA’s {\it NICER}, enormous studies on their implications on the neutron star EOS have been performed.
However, the $M$ and $R$ results from the {\it NICER} team are assuming neutron star EOS. They cannot be used to constrain the quark star EOS. See also discussions in~\citet{2020PhRvD.101d3003Z}.

After decades of speculation of self-bound strange quark matter being the physical nature of pulsar-like objects~\citep{
1971PhRvD...4.1601B,Terazawa1979,1984PhRvD..30..272W}, namely all pulsar-like objects should be quark stars, the idea is still neither proved nor excluded. See also discussions in~\citet{1985PhLB..160..181B,1990MPLA....5.2197G,2005PrPNP..54..193W}.
The quark stars' EOS remains implicit but frequently discussed as a viable alternative to the neutron star models. 
We here perform a Bayesian analysis to study the quark star EOS using the {\it NICER} PSR J0030+0451 data~\citep{2019ApJ...887L..24M,2019ApJ...887L..21R}.
As far as we know, such an analysis has not been performed elsewhere.
Since it is not clear whether quark stars have a crust or not, in this exploratory work we assume pulsar PSR J0030+0451 has a similar low-density crust as neutron stars, for example, a normal floating crust supported by electrostatic forces~\citep[e.g.,][]{2010MNRAS.402.2715L,2017ApJ...837...81W,2018RAA....18...82W}.
Such an assumption is necessary because bare quark stars cannot effectively generate surface thermal X-ray radiation, and the {\it NICER} estimation of mass and radius is based on standard nuclear models for the crust.
Although the crust effects on the star's mass and radius 
should be very limited
~\citep[see our previous calculations in, e.g.,][]{2020JHEAp..28...19L}, detailed discussions are provided due to its crucial relevance of the present analysis.

\begin{figure*}
    \begin{minipage}[t]{0.95\linewidth}
    \includegraphics[width=3.2in]{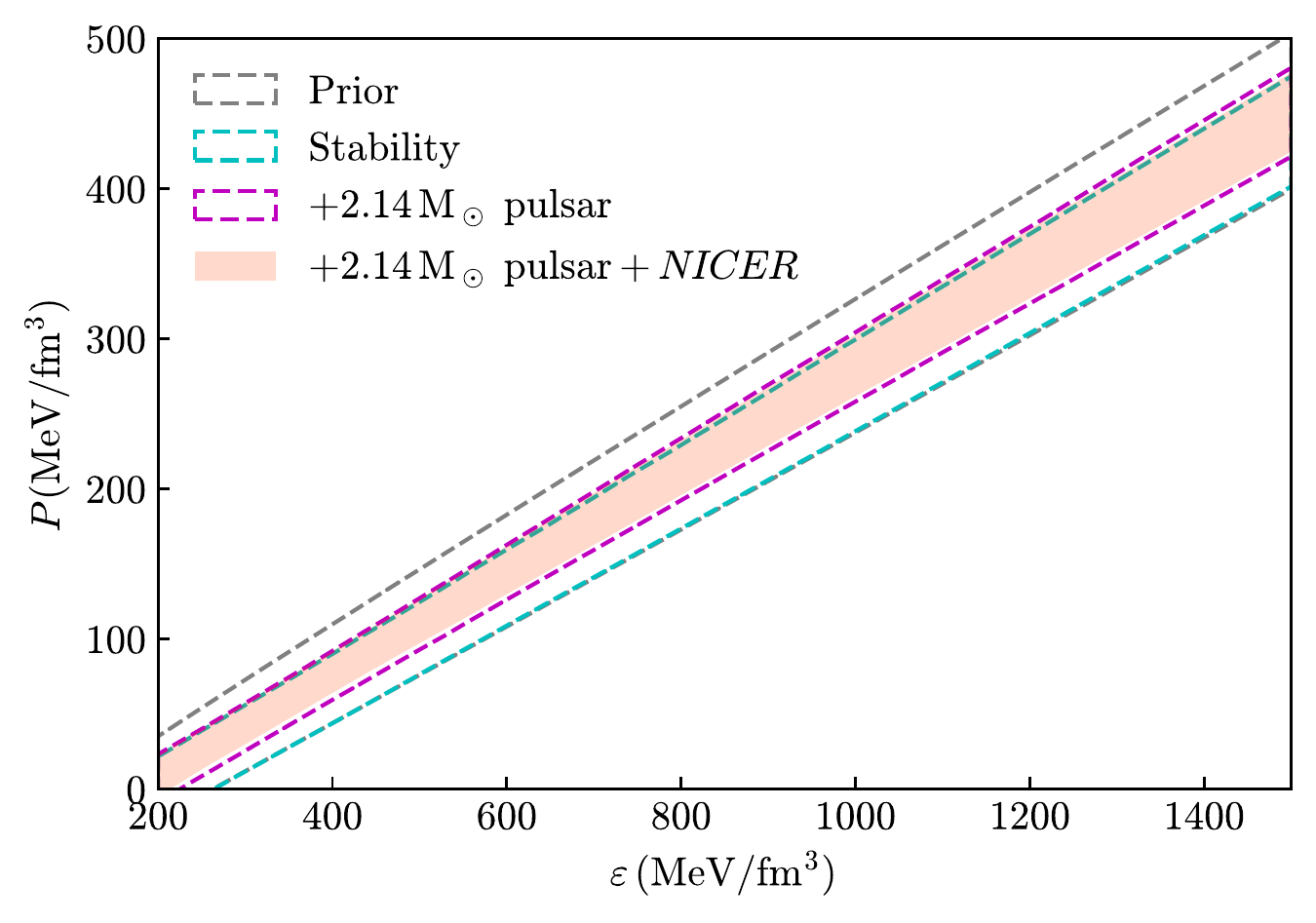}
    \includegraphics[width=3.2in]{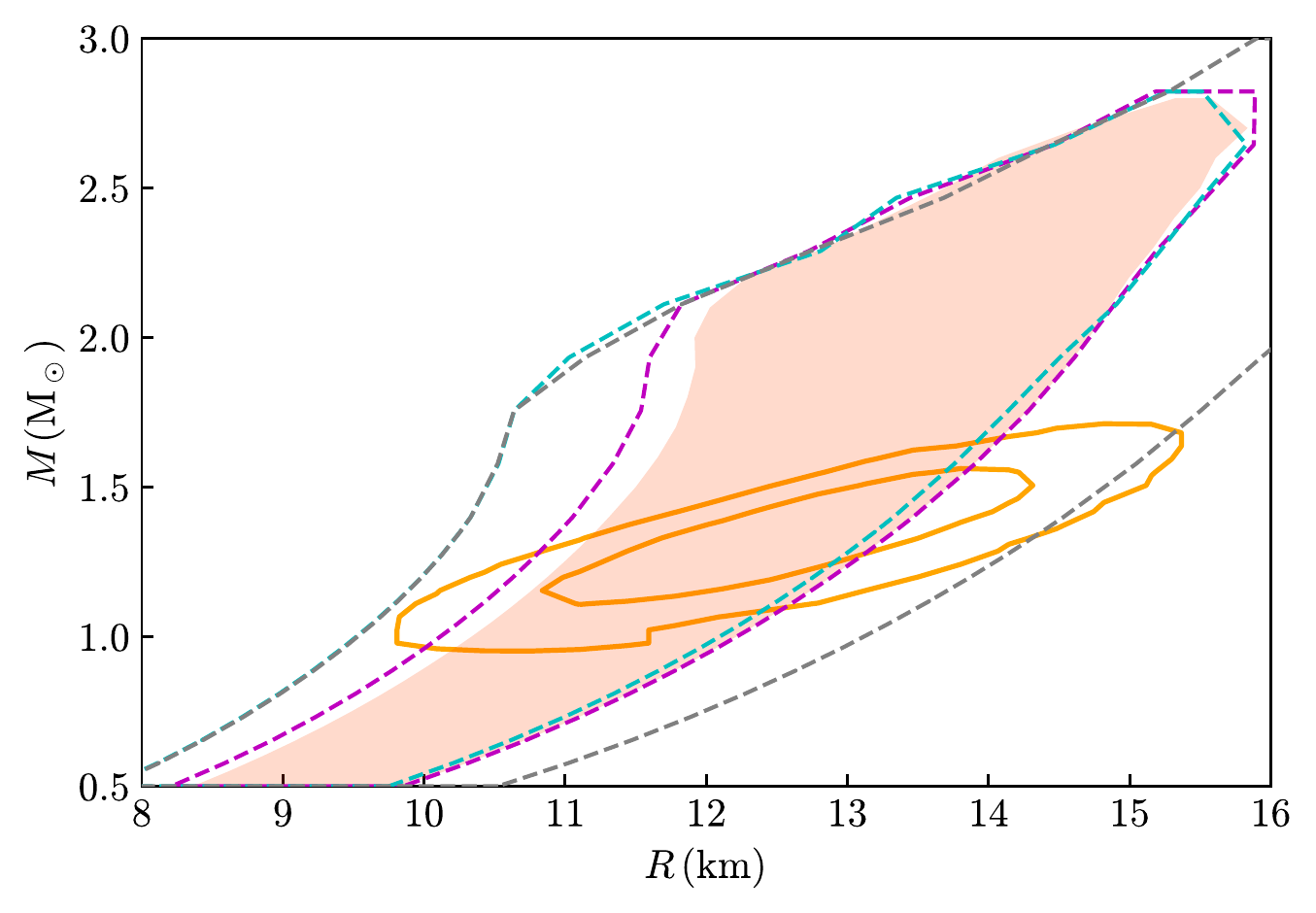}
    \end{minipage}
    \begin{minipage}[t]{0.95\linewidth}
    \includegraphics[width=3.2in]{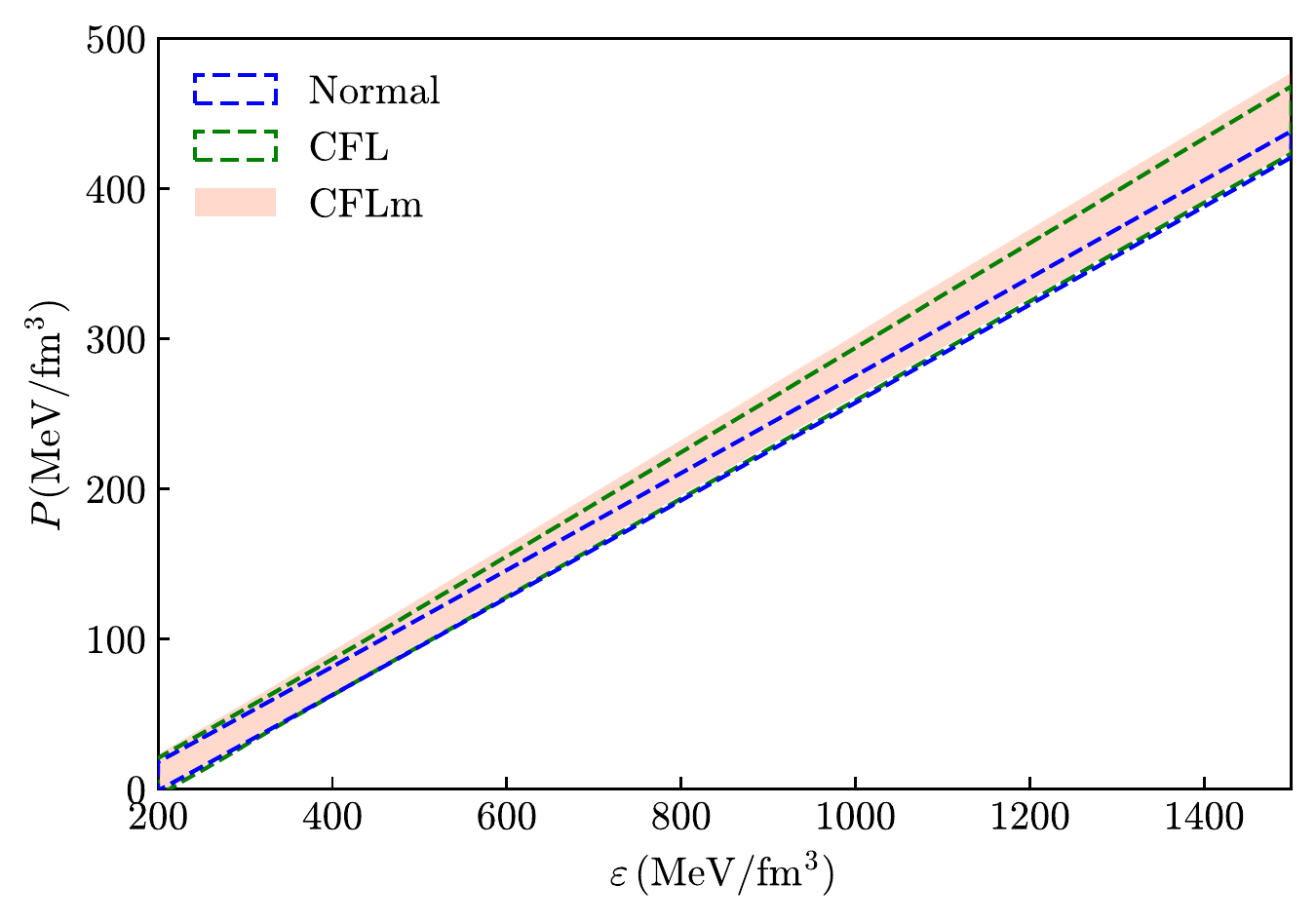}
    \includegraphics[width=3.2in]{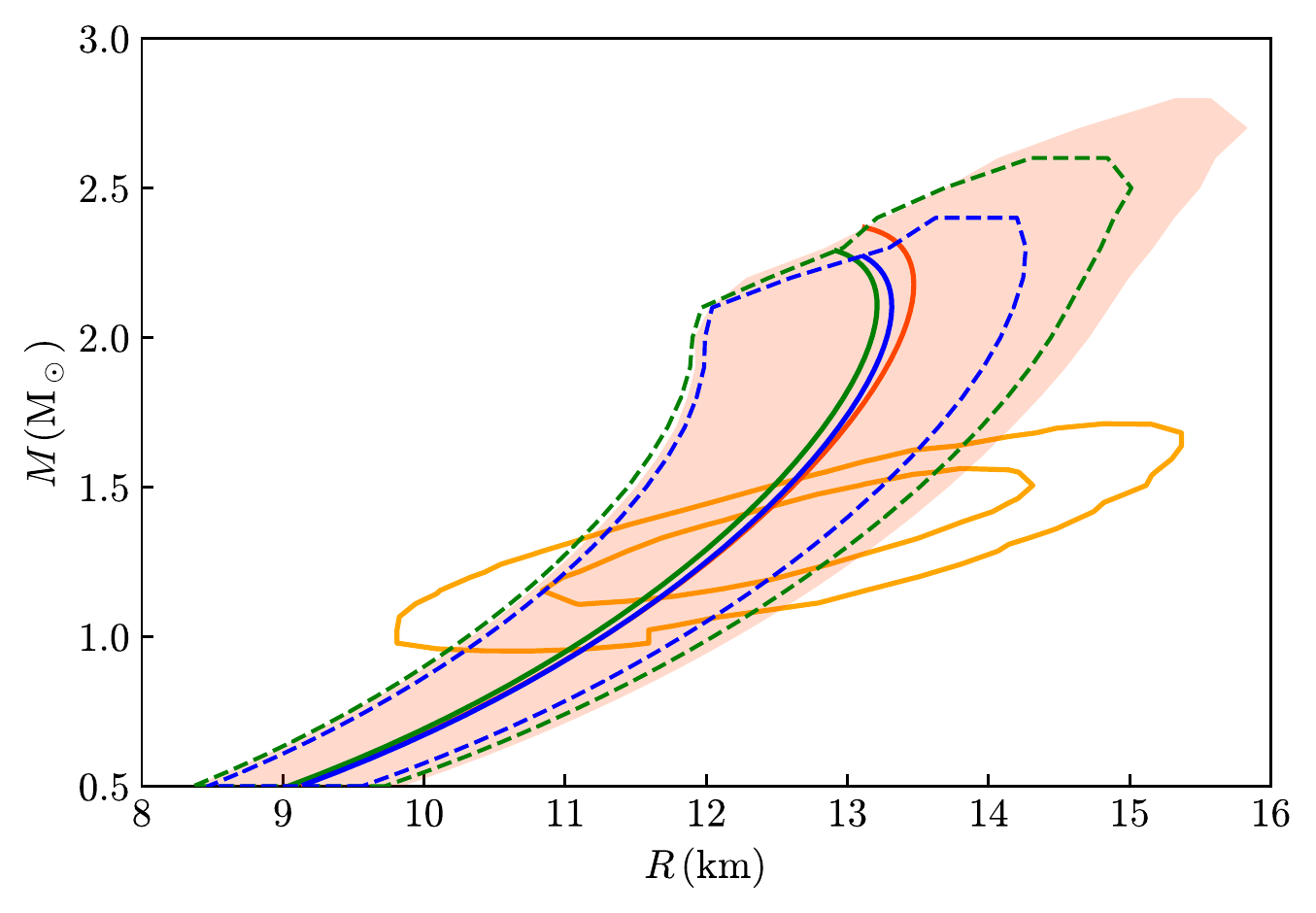}
    \end{minipage}
    \caption{Posterior distributions ($95\%$ confidence level) of the quark star EOS (left-hand panels) and the corresponding mass-radius relations (right-hand panels): The separate analysis of the stability test, the MSP J0740+6620 test, together with the prior and the joint MSP J0740+6620 \& PSR J0030+0451 test are shown in the upper panels for the four-parameter CFLm model; The results of the joint analysis within CFLm are compared to those of the Normal and CFL models in the lower panels; Shown together in the right-hand panels are the results of three exemplary quark star EOSs within the CFLm model ($B_{\rm eff},a_4,\Delta,m_s$): $(142.1,0.65,32.8,60), (136.9,0.55,37.3,50), (134.4,0.53,33.9,51)$ from Table~\ref{tb:1}.
    The orange contours in right panels indicate the 68\% and 95\% highest density posterior credible regions of mass-radius measurement results of \citet{riley_data}, respectively.
    }
\label{fig:1}
\end{figure*}

\begin{table}
  \centering
  \caption{Most probable intervals of the EOS parameters ($90\%$ confidence level) in three types of bag models (Normal \& CFL \& CFLm) constrained by the joint MSP J0740+6620 \& PSR J0030+0451 analysis (see details in Sec.~\ref{sect:obs}).
  }
\setlength{\tabcolsep}{9.pt}
\renewcommand\arraystretch{1.3}
  \begin{tabular}{lccc}    \hline \hline
   Parameters& & Prior 
   & $+2.14\Msun+${\it NICER}   \\
    \hline 
    $B_{\rm eff}^{1/4}/{\rm MeV}$& Normal & U$(125,150)$ 
    &  $130.3_{-4.7}^{+6.7}$  \\
    &CFL & U$(125,150)$ 
    &  $133.1_{-7.1}^{+11.7}$  \\
    & CFLm  & U$(125,150)$ 
    & $134.4_{-7.7}^{+11.1}$  \\
    \hline
    $a_4$ & Normal & U$(0.4,1)$ 
    & $0.58_{-0.09}^{+0.15}$  \\
    & CFL & U$(0.4,1)$ 
    &  $0.53_{-0.11}^{+0.16}$  \\
    & CFLm  & U$(0.4,1)$ 
    & $0.53_{-0.11}^{+0.18}$  \\
    \hline
    $\Delta/{\rm MeV}$
    & CFL& U$(0,100)$ 
    &  $35.1_{-30.6}^{+44.6}$  \\
    & CFLm  & U$(0,100)$ 
    & $33.9_{-30.6}^{+44.8}$  \\
    \hline
    $m_s/{\rm MeV}$ 
    & CFLm  & U$(0,150)$ 
    & $51_{-46}^{+66}$  \\
    \hline \hline
\end{tabular}
  \label{tb:1}
   \vspace{-0.2cm}
\end{table}

The paper is organized as follows. Section 2 is a brief overview of the popular bag-model description adopted for quark star EOS, where we consider both nonsuperfluid normal quark matter and superfluid matter in the Color-Flavor Locked (CFL) state;
Section 3 presents the employed {\it NICER} observations and the Bayesian analysis for the EOS; 
in Sec. 4, we discuss the quark star properties.
We then summarize the paper in Sec.~5. 

\begin{figure*}
    \begin{minipage}[t]{0.95\linewidth}
    \includegraphics[width=1.64in]{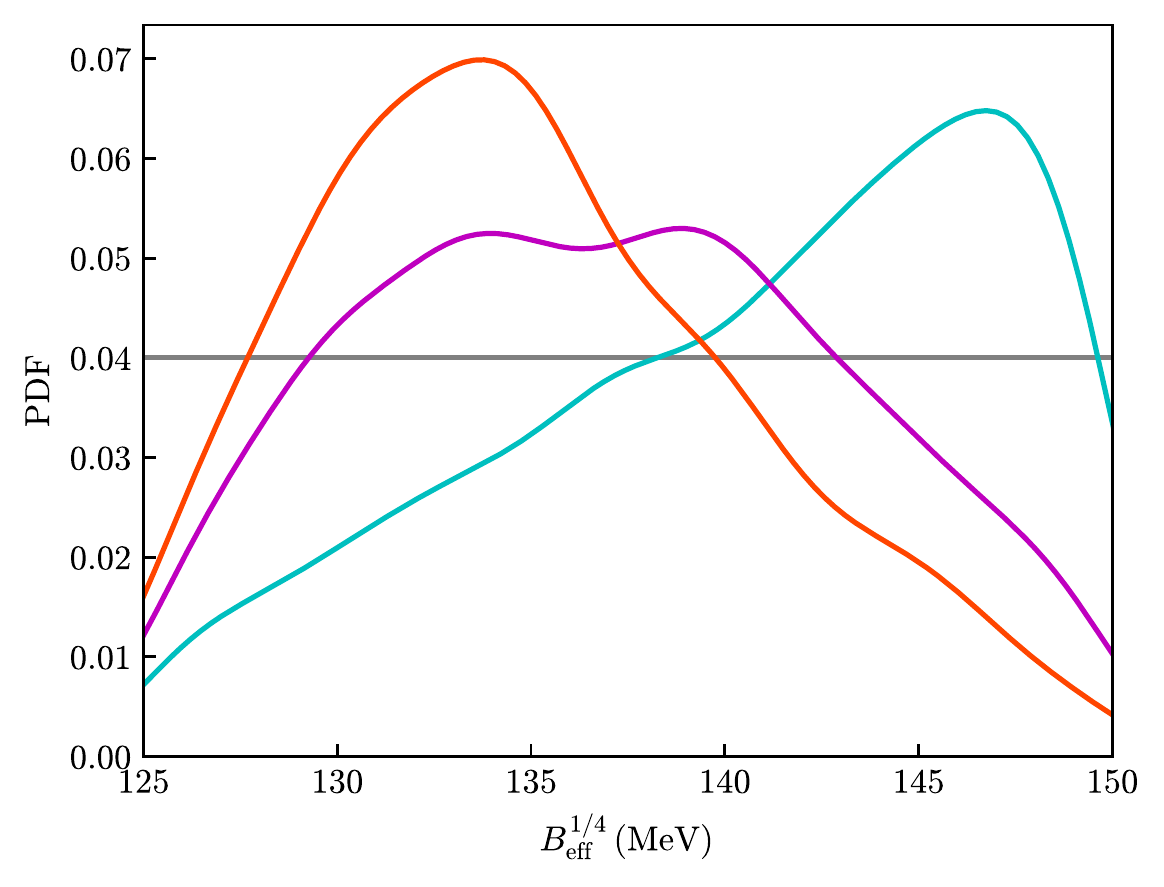}
    \includegraphics[width=1.64in]{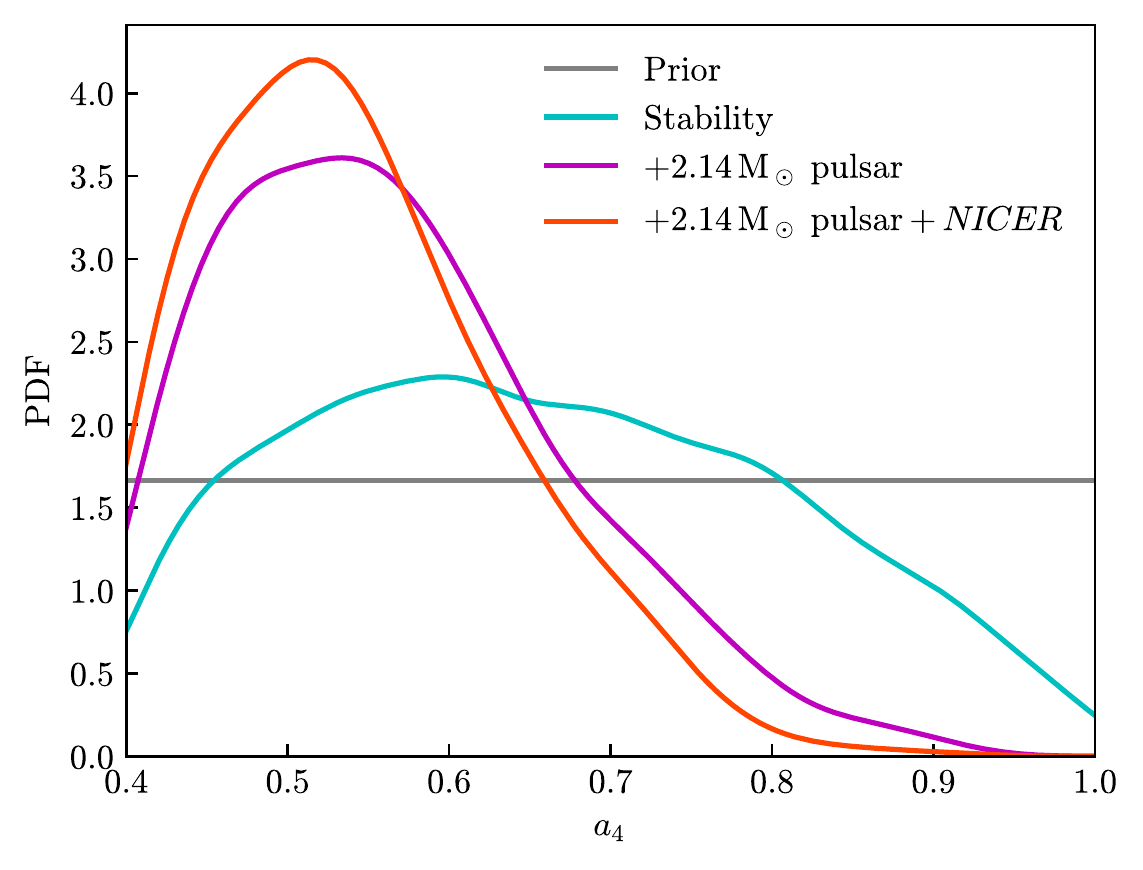}
    \includegraphics[width=1.64in]{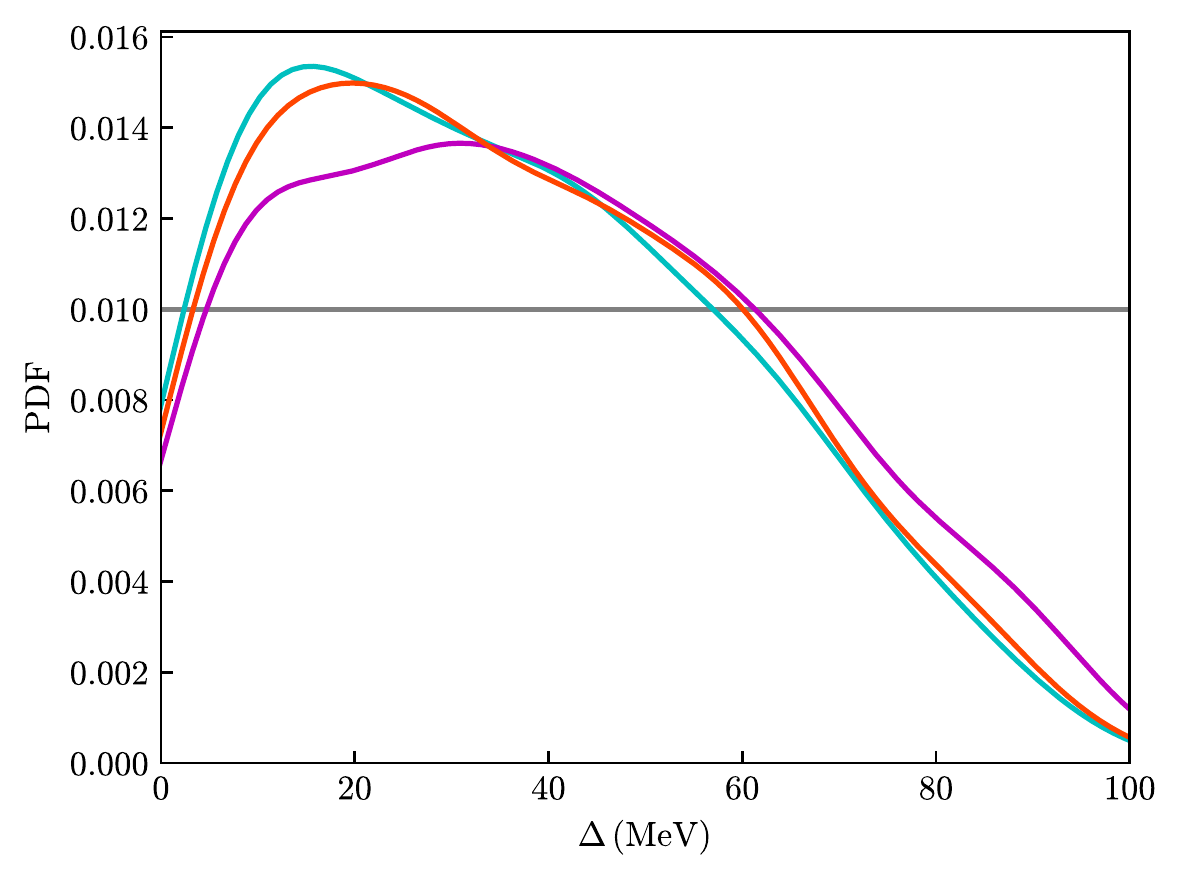}
    \includegraphics[width=1.64in]{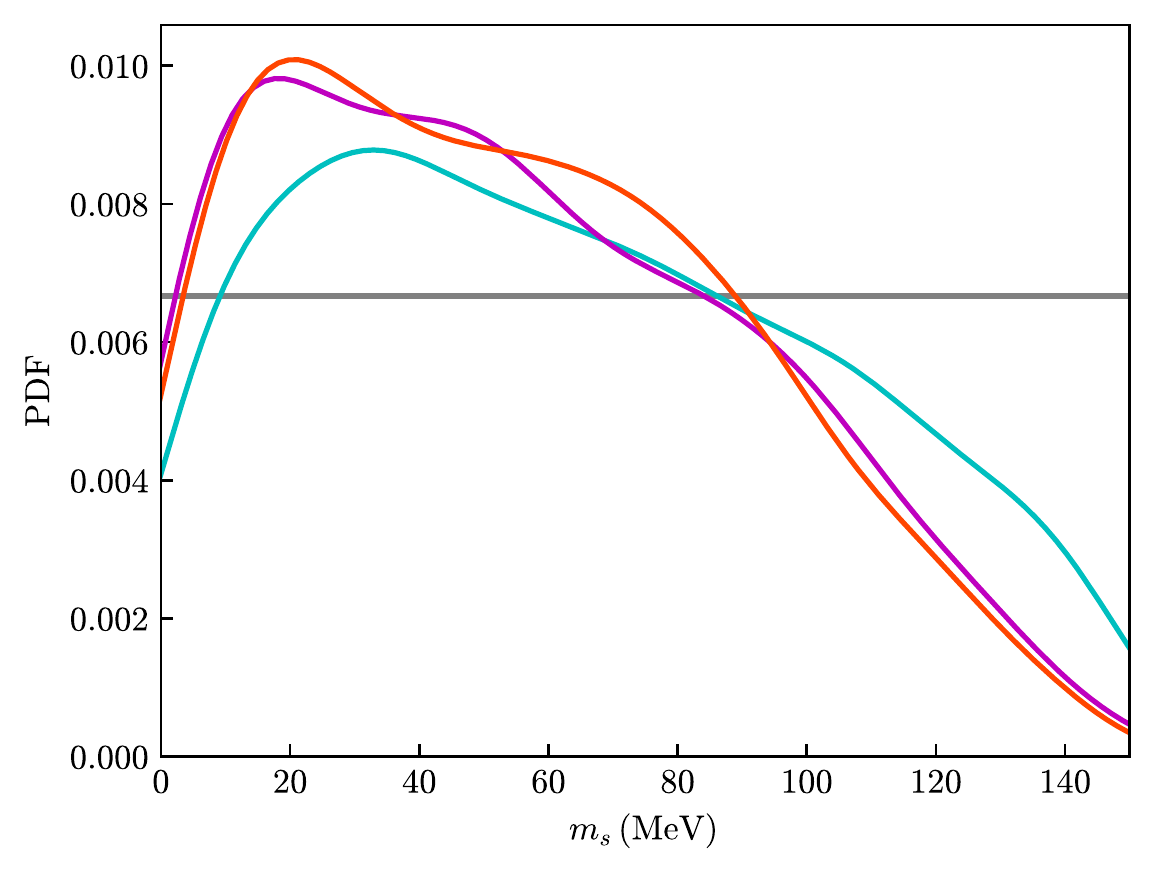}
    \end{minipage}
        \begin{minipage}[t]{0.95\linewidth}
    \includegraphics[width=2.3in]{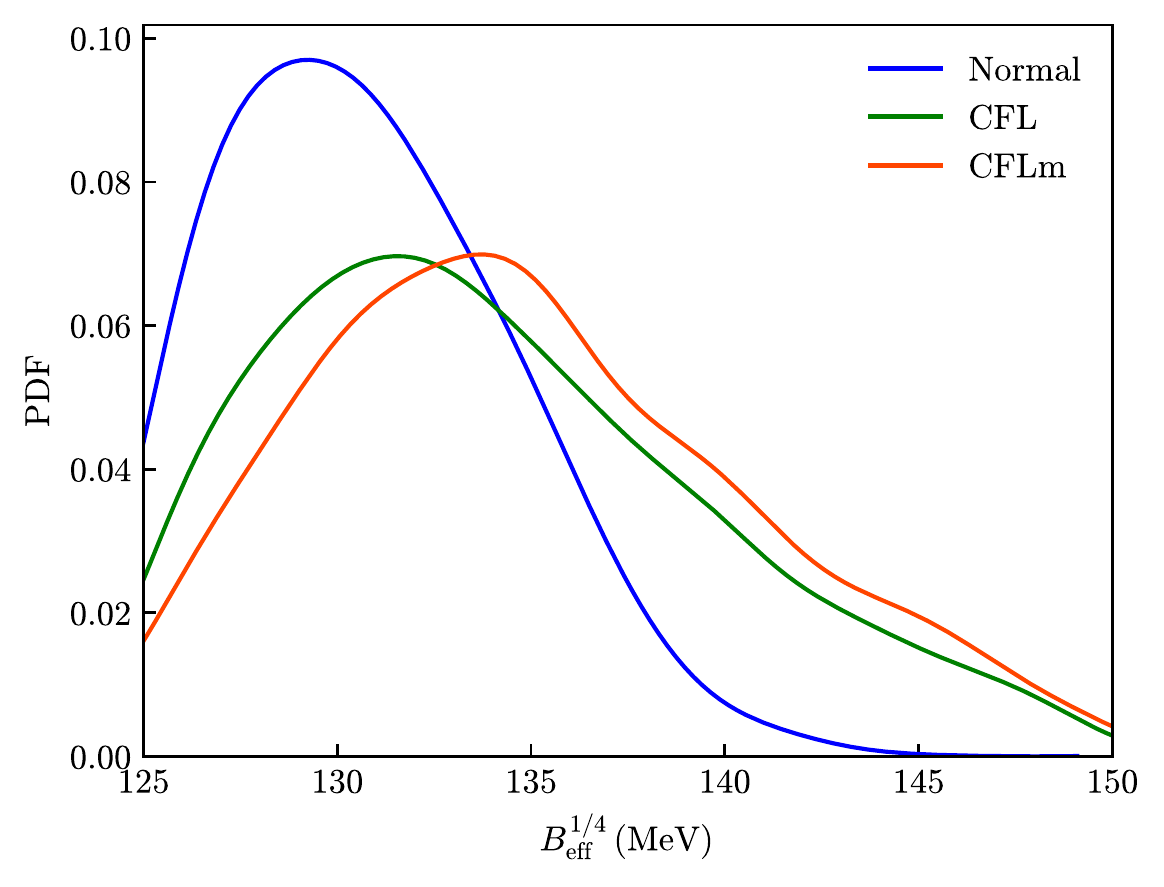}
    \includegraphics[width=2.3in]{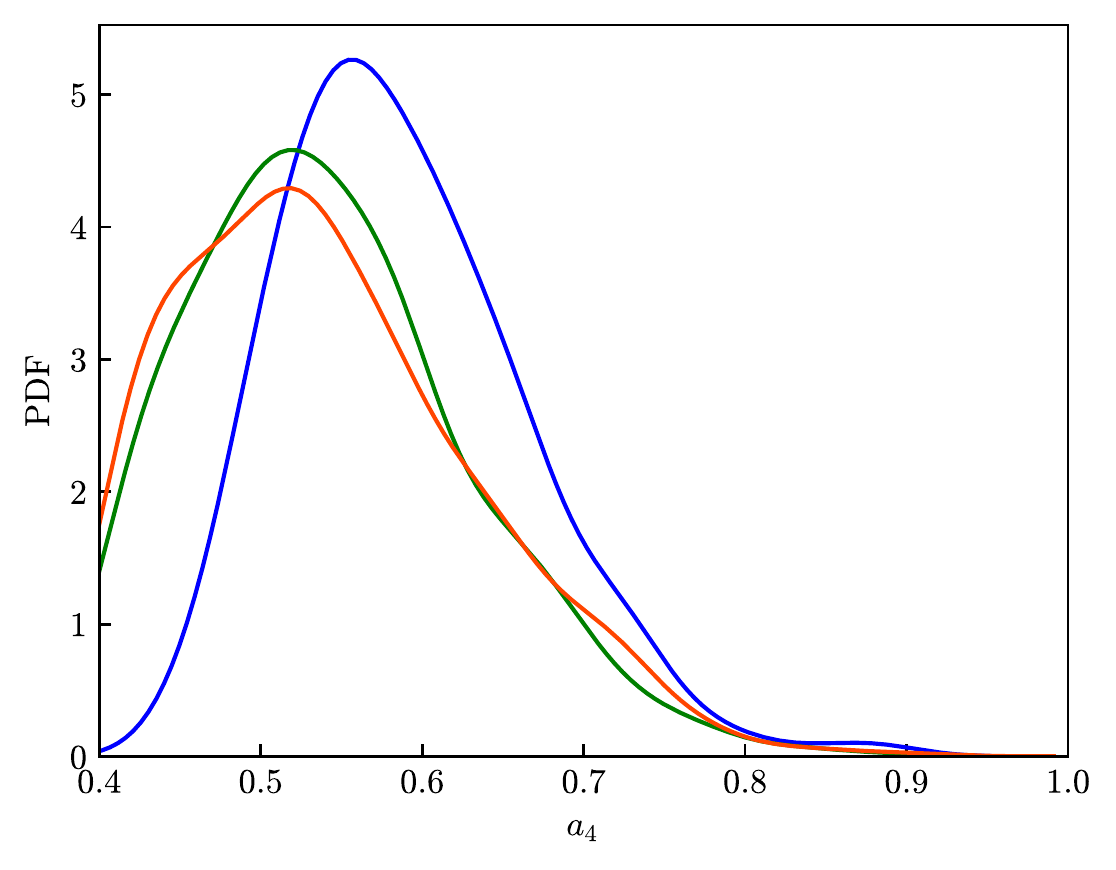}
    \includegraphics[width=2.3in]{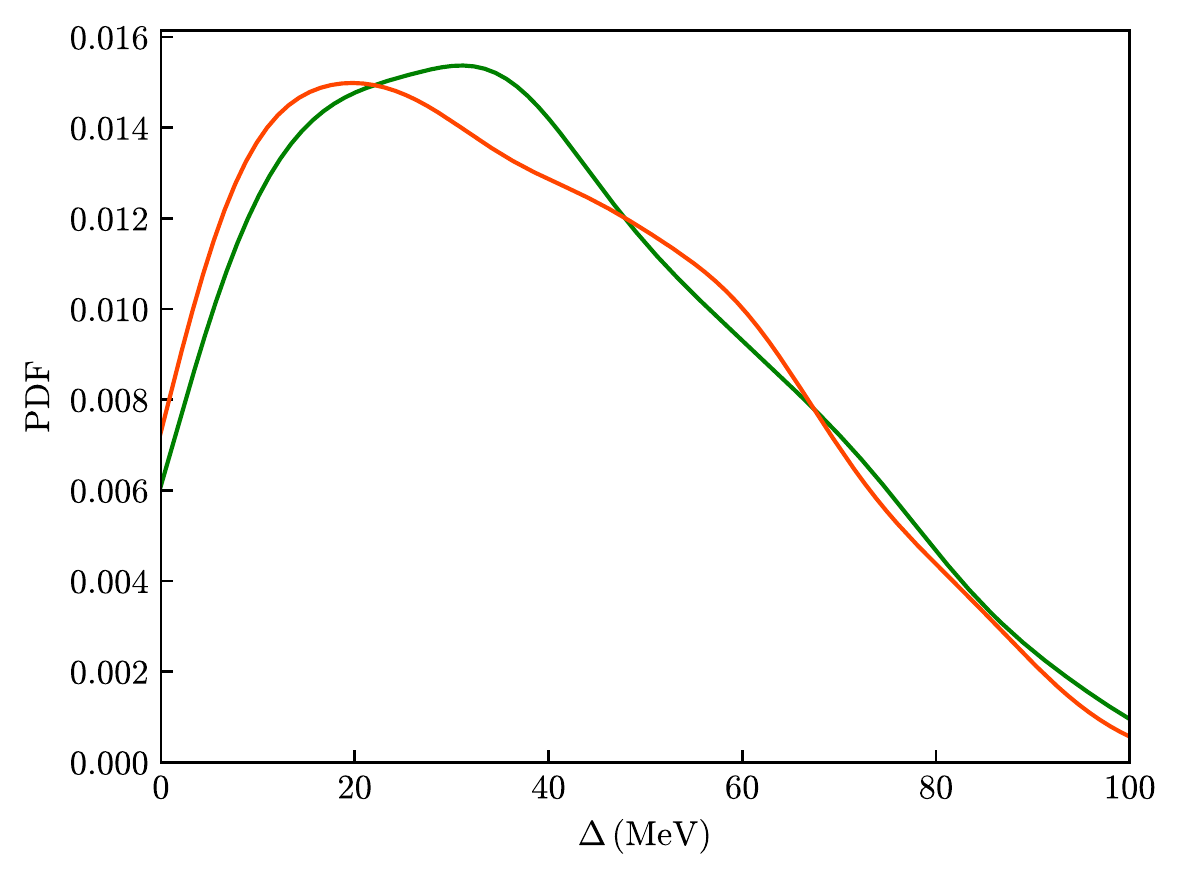}
    \end{minipage}
    \caption{Posterior PDFs of the quark star EOS parameters ($B_{\rm eff}^{1/4}, a_4, \Delta,m_s$) (90\% confidence level) in three types of bag models (Normal \& CFL \& CFLm) (see details in Sec.~\ref{sect:mit}) conditioned on the prior of three different analyses (see details in Sec.~\ref{sect:obs}). 
    The separate analysis of the stability test, the MSP J0740+6620 test, together with the prior and the joint MSP J0740+6620
     \& PSR J0030+0451 test are shown in the upper panel for the four-parameter CFLm model.
    The results of the joint analysis within CFLm are compared to those of the Normal and CFL models in the lower panel.}
    \label{fig:2}
\end{figure*}

\section{The bag models for quark star EOS}
\label{sect:mit}

For normal quark matter, the bag-model expressions for the grand canonical potential per unit volume are written as~\citep[e.g.,][]{1986ApJ...310..261A,1986A&A...160..121H,2005ApJ...629..969A,2016MNRAS.457.3101B,2017ApJ...844...41L,2018PhRvD..97h3015Z}:
\begin{equation}
\Omega_{\rm Normal}=\sum_{{i=u,\ d,\ s,\ e^-}}\Omega_i^0+\frac{3(1-a_4)}{4\pi^2}\mu^4+B_\mathrm{eff}~. \label{eq:canonicalpotential1}
\end{equation}
with the total baryon number density $n = (n_u + n_d + n_s)/3$ from $n_i = -(\partial \Omega / \partial \mu_i)_{\rm V}$. $\Omega_i^0$ is the grand canonical potential for particle type $i$ described as ideal Fermi gas. $\mu=(\mu_u+\mu_d+\mu_s)/3$ is the average quark chemical potential. 
$B_{\rm eff}$ accounts for the contributions from the QCD vacuum, and $a_4$ characterizes the perturbative QCD contribution from one-gluon exchange for gluon interaction.
If quark matter is in the CFL phase, an additional term corresponding to the pairing energy has to be added~\citep{2001PhRvD..64g4017A,2001PhRvL..86.3492R,2002PhRvD..66g4017L},
\begin{equation}
    \Omega_{\rm CFL}=\Omega_{\rm Normal}+\frac{3m_s^4-48\Delta^2\mu^2}{16\pi^2}\ ,
\end{equation}
where $\Delta$ is the uncertain CFL pairing gap, which can be as high as $100$ MeV.

In the following, we consider the strange quark matter constituting the stars with three types of models in the bag-model description, namely: \\
(i) two-parameter model \textbf{Normal($B_{\rm eff},a_4$)}, where we consider normal quark matter. Also u and d quarks are regarded as massless ($m_u=m_d=0$) while fixing the strange quark mass $m_s=100~\rm MeV$;\\
(ii) three-parameter model \textbf{CFL($B_{\rm eff},a_4,\Delta$)}, where we consider CFL superfluid quark matter: As in (i), $m_u=m_d=0,m_s=100~\rm MeV$;\\
(iii) four-parameter model \textbf{CFLm($B_{\rm eff},a_4,\Delta,m_s$)}, where we vary the strange quark mass $m_s$ in the CFL superfluid quark matter in (ii) in the range of $0$ to $150~\rm MeV$.

\begin{table*}
  \centering
  \caption{Most probable intervals of three quark star properties ($90\%$ confidence level) in three types of bag models (Normal \& CFL \& CFLm) (see details in Sec.~\ref{sect:mit}) conditioned on the prior and three different data sets (see details in Sec.~\ref{sect:obs}). $M_{\rm TOV}$ is the maximum mass. $R_{\rm 1.4}$ is the radius of a $1.4\Msun$ star. $n_{\rm surf}$ is the surface density.
  }
\setlength{\tabcolsep}{0.9pt}
\renewcommand\arraystretch{1.7}
\begin{tabular*}{\hsize}{@{}@{\extracolsep{\fill}}lccccc@{}}\hline\hline
   Parameters& & Prior & Stability &  $+2.14 \Msun$ pulsar  & $+2.14\Msun+${\it NICER}   \\
    \hline
    $M_{\rm TOV}/ \Msun$& Normal & $2.10_{-0.29}^{+0.37}$ & $2.05_{-0.22}^{+0.37}$ &  $2.27_{-0.21}^{+0.19}$&  $2.31_{-0.19}^{+0.15}$  \\
    &CFL & $2.29_{-0.42}^{+0.66}$ & $2.09_{-0.24}^{+0.37}$ &  $2.29_{-0.23}^{+0.23}$ &  $2.34_{-0.21}^{+0.20}$  \\
    & CFLm  & $2.33_{-0.44}^{+0.69}$ & $2.15_{-0.28}^{+0.43}$ & $2.32_{-0.24}^{+0.31}$ & $2.38_{-0.23}^{+0.26}$  \\
    \hline
    $R_{1.4}/\rm km$ & Normal & $11.48_{-1.28}^{+1.52}$  & $11.26_{-0.99}^{+1.57}$ &  $12.20_{-0.91}^{+0.77}$ & $12.38_{-0.82}^{+0.62}$  \\
    & CFL   & $12.03_{-1.58}^{+2.01}$ & $11.30_{-0.95}^{+1.55}$ &  $12.07_{-0.95}^{+0.99}$  &  $12.32_{-0.90}^{+0.81}$  \\
    & CFLm  & $12.13_{-1.62}^{+2.06}$ & $11.48_{-1.04}^{+1.58}$ &  $12.11_{-0.90}^{+1.18}$ & $12.39_{-0.94}^{+0.93}$  \\
    \hline
    $n_{\rm surf}/\rm fm^{-3}$ & Normal& $0.22_{-0.05}^{+0.07}$  & $0.24_{-0.07}^{+0.06}$ &  $0.19_{-0.03}^{+0.05}$ & $0.18_{-0.02}^{+0.04}$  \\
    & CFL& $0.21_{-0.05}^{+0.07}$ & $0.24_{-0.07}^{+0.06}$ &  $0.20_{-0.04}^{+0.05}$ &  $0.19_{-0.03}^{+0.04}$  \\
    & CFLm  & $0.21_{-0.05}^{+0.07}$ & $0.23_{-0.07}^{+0.06}$ &  $0.20_{-0.04}^{+0.05}$ & $0.19_{-0.03}^{+0.05}$  \\
    \hline \hline
\end{tabular*}
  \label{tb:2}
\end{table*}

\begin{figure*}
    \includegraphics[width=7.in]{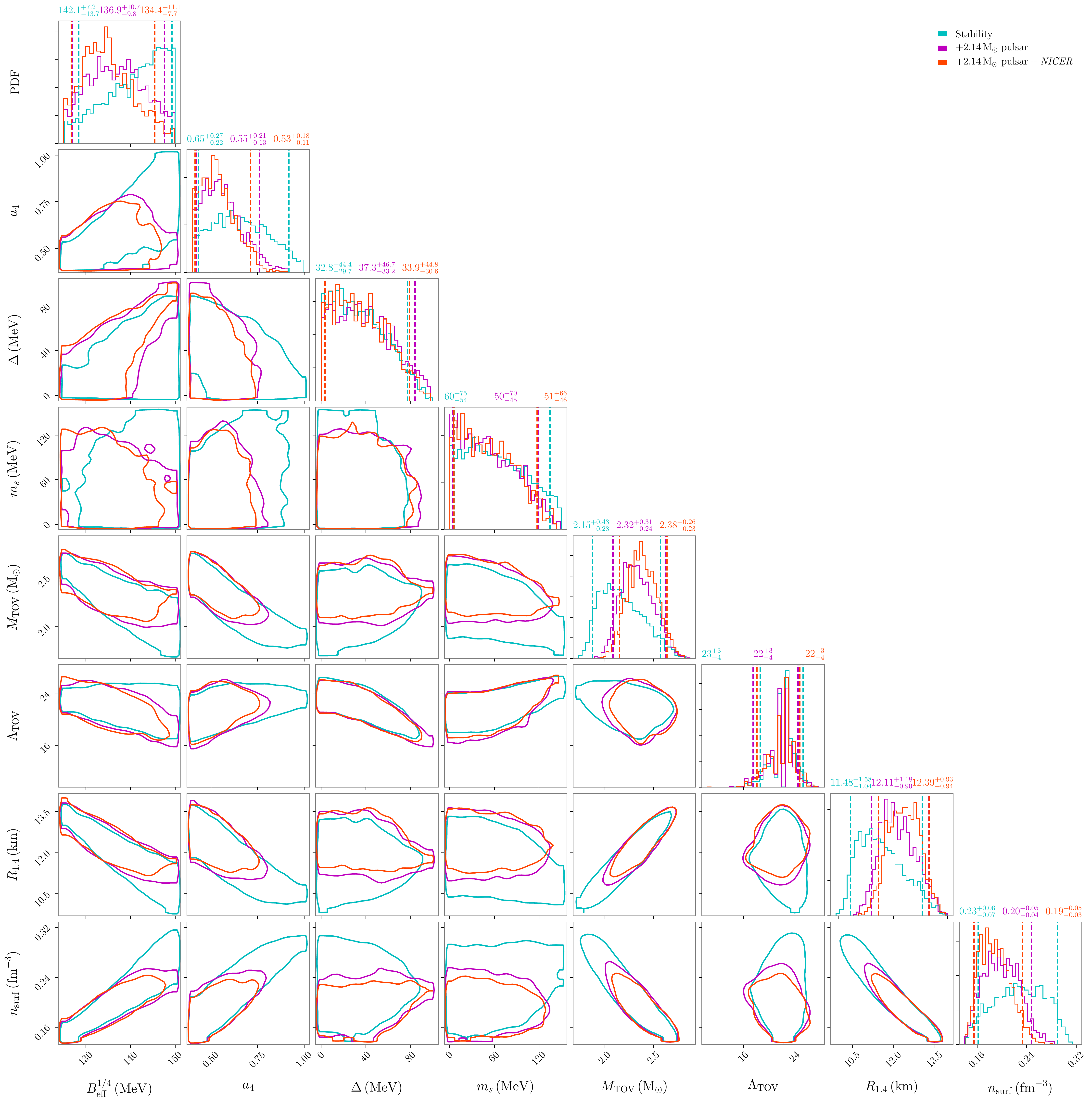}
    \caption{Posteriors distributions of the quark star EOS parameters ($B_{\rm eff}^{1/4}, a_4, \Delta,m_s$) and quark star properties ($M_{\rm TOV}, \Lambda_{\rm TOV}, R_{\rm 1.4},n_{\rm surf}$), together with those of the stability conditions and/or the constraints from the $2.14\Msun$ pulsar.
    The contours are the $90\%$ credible regions for the parameters.
    The cyan, magenta, and red contours represent the results conditioned on the uniform prior for the stability test, the MSP J0740+6620 test, and the joint MSP J0740+6620 and PSR J0030+0451 test, respectively (see Sec.~\ref{sect:obs} for details).
     }
   \label{fig:3}
\end{figure*}

\section{Stability, Constraint, and Bayesian analysis}
\label{sect:obs}

We employ Bayesian analysis to estimate the bag model parameters ($B_{\rm eff},a_4,\Delta,m_s$) and deduce the masses and radii of quark stars. According to the Bayes' theorem, the posterior distributions of the model parameters $\boldsymbol\theta$ can be written as
\begin{equation}\label{eq:bayes theorem}
    p(\boldsymbol\theta|\boldsymbol d)\propto p(\boldsymbol\theta)p(\boldsymbol d|\boldsymbol\theta)
\end{equation}
where $\boldsymbol d$ is the observational data set.

$p(\boldsymbol\theta)$ in Eq.(\ref{eq:bayes theorem}) is the prior which reflects our preliminary knowledge of the model parameters. We choose a uniform distribution of $B_{\rm eff}^{1/4}$/$a_4$ in the range of $[125,150]\,{\rm MeV}$/$[0.4,1]$ according to our previous work~\citep{2018PhRvD..97h3015Z}. 
We also assign a reasonably wide boundary for the pairing gap/strange quark mass as $\Delta \in[0, 100]\,{\rm MeV}$/$m_s \in[0,150]\,{\rm MeV}$ with a uniform distribution. 

$p(\boldsymbol d|\boldsymbol\theta)$ in Eq.(\ref{eq:bayes theorem}) is the likelihood of observational data $\boldsymbol d$ given parameter set $\boldsymbol\theta$. In the present analysis, we will consider three constraints from both theoretical consideration and practical observation, which are explained as follows.

{\bf Stability arguments.} Before the input of any observational data, two stability constraints for the quark star EOS are adopted: First, the energy per baryon for non-strange quark matter should satisfy $(E/A)_{\rm ud}\geq934\,{\rm MeV}$ to guarantee the observed stability of atomic nuclei; Secondly, $(E/A)_{\rm uds}\leq930\,{\rm MeV}$ is required, according to the hypothesis that the strange quark matter is absolutely stable~\citep{
,1971PhRvD...4.1601B,Terazawa1979,1984PhRvD..30..272W}. Therefore, the likelihood contribution of the stability conditions is expressed as
\begin{equation}
    p((E/A)_{\rm ud},(E/A)_{\rm uds}|\boldsymbol\theta_{\rm EOS})=  \left\{%
  \begin{array}{l}
  1,\quad (E/A)_{\rm ud}\geq 934\,{\rm MeV }\,\&\\(E/A)_{\rm uds}\leq 930\,{\rm MeV},\\
  0,\quad {\rm others}.
  \end{array}
  \right.
\end{equation}

{\bf Constraints from MSP J0740+6620.} To ensure the EOS is stiff enough to support the presently known heaviest pulsars,
we adopt the mass measurement of MSP J0740+6620 detected through Shapiro delay~\citep{2020NatAs...4...72C}, $M=2.14^{+0.10}_{-0.09} \Msun$ ($68.3\%$ credibility interval), to set a lower limit on the quark star maximum mass $M_{\rm TOV}$. In practice, we sample a mass $M_0$ from the distribution of this source in each MCMC iteration step and require the quark star maximum mass should be larger than this mass, i.e., $M_{\rm TOV}\geq M_0$. Those EOS parameter sets that cannot support such a mass $M_0$ will be rejected in MCMC sampling. The likelihood then reads
\begin{equation}
    p(M_0|\boldsymbol \theta_{\rm EOS})=\left\{%
  \begin{array}{l}
  1,\quad M_{\rm TOV}(\boldsymbol \theta_{\rm EOS})\geq M_0,\\
  0,\quad M_{\rm TOV}(\boldsymbol \theta_{\rm EOS})<M_0.
  \end{array}
  \right.
\end{equation}

{\bf {\it NICER} PSR J0030+0451 data.} Here we incorporate the recent simultaneous measurements of mass and radius of PSR J0030+0451 from {\it NICER}~\citep{2019ApJ...887L..24M,2019ApJ...887L..21R}. Since the two results of~\citet{2019ApJ...887L..21R} and \citet{2019ApJ...887L..24M} are consistent with each other, we only adopt the best-fitting scenario within the ST+PST model of \citet{2019ApJ...887L..21R} for the present analysis. In this case, we need an extra parameter, the central energy density $\varepsilon_c$, because different central energy densities correspond to different masses and radii. Thus, by using a Gaussian Kernel Density Estimation (KDE) of the posterior samples $\vec{S}$ of the mass and radius given by \citet{riley_data}, the likelihood function can be expressed as 
\begin{equation}
    p(M, R|\boldsymbol\theta_{\rm EOS},\varepsilon_c) = {\rm KDE}(M, R\mid \vec{S}),
\end{equation}
where $(M, R)$ are obtained by solving the TOV equation with EOS set $\boldsymbol\theta_{\rm EOS}$ and central density $\varepsilon_c$.

\section{Results and discussion}\label{sec:results}
\label{sect:result}

\subsection{Quark star EOS from three types of bag-model parametrization}

We first show in Fig.~\ref{fig:1} the posterior distributions of the quark star EOSs and the corresponding mass-radius relations.
The results of three different analyses (see details in Sec.~\ref{sect:obs}) within CFLm are compared in the upper panels.
We see that the stability arguments which ensure that quark stars constitute self-bound three-flavoured quark matter exclude too stiff EOSs and consequently disfavour superheavy quark stars above $\sim2.6\Msun$, to the $95\%$ confidence level.
The $2.14\Msun$ pulsar data, on the other hand, exclude some parameter space of soft EOSs.
Adding the {\it NICER} data excludes further a small parameter space of soft EOSs and constrains the EOS in the pink shaded region.
Compared to the $\it NICER$ analysis adopting the neutron star models, our analysis based on quark star models results in a relatively more compact star for a certain mass.
In the lower panels of the same figure, we compare the results of three employed EOS parametrization in the bag model.
It is evident that the allowed EOS parameter space is enlarged, mostly in stiff EOS cases, when we add one free parameter in the model. The detailed results of the parameter ranges are collected in Table \ref{tb:1}.

To understand better the resulting parameter ranges of the quark star EOS, we show in Fig. \ref{fig:2} the posterior probability distribution functions (PDFs) of the EOS parameters for various bag models introduced in Sec. \ref{sect:mit}.
As seen in Fig. \ref{fig:2}, from the stability arguments, large/small values of $B_{\rm eff}$ are not allowed by the requirement $uds$/$ud$ matter is stable/unstable in bulk~\citep{1984PhRvD..30.2379F}. 
Also, the EOS should be stiff enough to support a heavy star of mass $2.14\Msun$, thus large $B_{\rm eff}$ values are further excluded due to the anticorrelation between $B_{\rm eff}$ and the EOS stiffness. 
For example, in the normal-matter case of non-interacting quarks with vanishing mass, there is a relation tells $M_{\rm TOV}\varpropto B_{\rm eff}^{-1/2}$~\citep{1984PhRvD..30..272W}. 
The {\it NICER} data tend to support even stiffer EOSs than the $2.14\Msun$ one, as can be seen more clearly in the following section of the resulting stellar properties. 

\subsection{Maximum mass and typical radius of quark stars}

The $90\%$ confidence boundaries of three quark star properties ($M_{\rm TOV}, R_{\rm 1.4}, n_{\rm surf}$) are reported in Table \ref{tb:2}.
For the two dominant model parameters ($B_{\rm eff}, a_4$) relevant to the quark matter stability, the previous Fig.~\ref{fig:2} has shown that their parameter ranges in the two-parameter normal matter model are different from those of three/four-parameter CFL matter models.
Nevertheless, three kinds of parametrization models all result in similar star properties, e.g., a maximum mass $M_{\rm TOV}\sim2.3\Msun$, a typical radius $R_{\rm 1.4}\sim12.3~\rm km$, a surface density $n_{\rm surf}\sim0.18\fm3$. 
Therefore the prior dependence is relatively modest as long as we adopt necessary constraints from nuclear physics and astrophysics.
In particular, the maximum mass of quark stars is $2.38^{+0.26}_{-0.23}\Msun$ for the four-parameter model, where the model flexibility from $\Delta$ and $m_s$ are both taken into account.
We mention again that, compared to the analysis based on the neutron star models~\citep{2019ApJ...887L..24M,2019ApJ...887L..21R}, the quark stars are more compact, $\sim12.3~\rm km$ vs. $\sim13~\rm km$.

Since the current Bayesian inference directly connects the astrophysical observables with the underlying quark star EOSs, we present further in Fig~\ref{fig:3} the marginalized posterior probability of all four bag model EOS parameters ($B_{\rm eff}^{1/4}, a_4, \Delta,m_s$) plus four quark star properties ($M_{\rm TOV}, \Lambda_{\rm TOV}, R_{\rm 1.4}, n_{\rm surf}$) and their correlations.
It is demonstrated that $M_{\rm TOV}/R_{\rm 1.4}$ anticorrelates with $B_{\rm eff}^{1/4}$ and even better with $a_4$, but not sensitive to $\Delta$ and $m_s$.
Although there is relatively good $\Delta-\Lambda_{\rm TOV}$ anticorrelation, no quark star observed properties depend sensitively on $m_s$. 
$m_s$ therefore cannot be well-constrained by the data, but it slightly affects both $M_{\rm TOV}$ and $R_{\rm TOV}$ due to its softening effects on the EOS~\citep{2018PhRvD..97h3015Z}.
The $\Delta-\Lambda_{\rm TOV}$ correlation revealed here, on the other hand, may shed light on the uncertain color superconductivity gap in future measurements of binary merger events, with a component mass close to the maximum mass.
In fact, the current analysis based on X-ray observations of pulsars can serve as independent constraints on the quark star EOSs and is ready to be confronted with the study from the gravitational wave signals. Those analyses on the GW170817~\citep{2017PhRvL.119p1101A} and GW190425~\citep{2020ApJ...892L...3A} data from LIGO/Virgo will be reported in a separate work~\citep{2021arXiv210713997M}.

\subsection{Crustal effects on the mass-radius relations of quark stars}

\begin{figure}
    \includegraphics[width=3.5in]{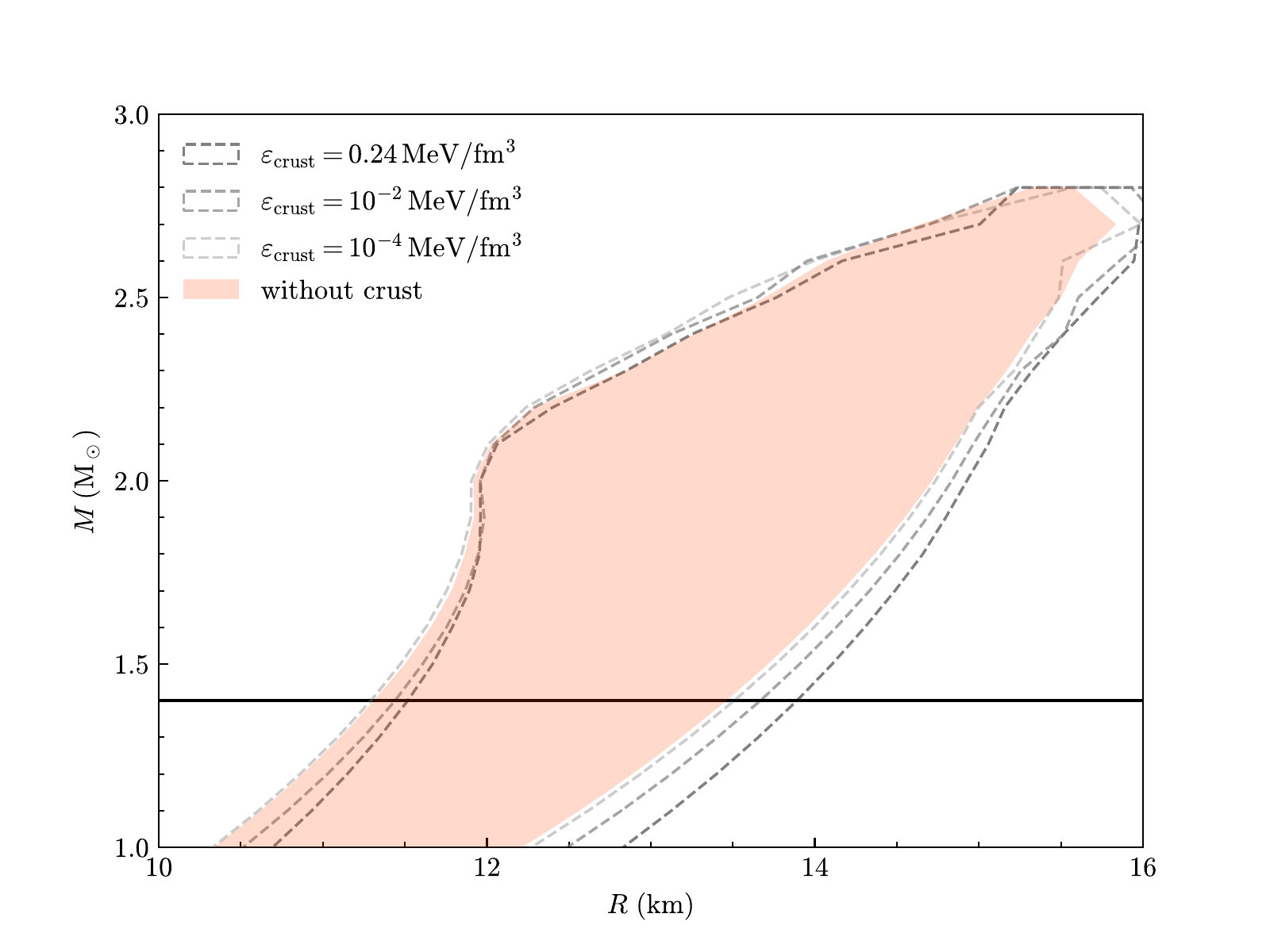}
    \caption{Posteriors distributions (95\% confidence level) of the mass-radius relation for quark stars with (grey-dashed lines) or without a crust (red-shaded region). The horizontal line corresponds to $M=1.4\Msun$.}
   \label{fig:4}
\end{figure}

In the discussion above, we consider quark stars without a crust. 
However, in the NICER estimation of mass and radius, a normal crust is necessary. 
The crust might add about $1$ km to $R_{\rm 1.4}$~\citep{2017A&A...599A.119Z} for neutron stars, which exceeds the design accuracy of the NICER experiment. 
The present section is devoted to the crust effects on the mass and radius of quark stars in our analysis.

Different to the neutron star case, where there is a layer of inner crust, and free neutrons are present between the neutron drip density $\varepsilon_{\rm drip}= 0.24\,{\rm MeV/fm^3}~ (4.3\times 10^{11} {\rm g/cm^3)}$ and the nuclear saturation density (i.e., the core part), quark stars may only have a thin nuclear (outer-)crust supported by an electric dipole layer, surrounding the quark matter core~\citep{2005PrPNP..54..193W}. Therefore, the maximum density of the quark star crust is limited by the neutron drip density, above which neutrons would gravitate toward the strange-quark matter core. 

Presently, we choose $\varepsilon_{\rm crust}$ to be the neutron drip density ($\varepsilon_{\rm crust}= 0.24\,{\rm MeV/fm^{3}}$) as well as two lower values of $10^{-2}\,{\rm MeV/fm^{3}}$ and $10^{-4}\,{\rm MeV/fm^{3}}$.
We then match the low-density nuclear EOS~\citep[using the standard one proposed by][]{1971ApJ...170..299B} to the high-density quark matter EOSs (using the representative CFLm model) at the three choices of crust density $\varepsilon_{\rm crust}$, and generate three EOS families for quark stars with a crust. 
We re-perform the joint MSP J0740+6620 and PSR J0030+0451 analysis for the three families of EOSs within CFLm. 

The inferred CFLm parameters (90\% confidence level) are reported in Table \ref{tb:3}. 
Comparing with our previous results listed in Table \ref{tb:1}, we see the inclusion of a crust yields model parameter differences at most 2-3\% at the largest maximal inner crust density $\varepsilon_{\rm crust}= \varepsilon_{\rm drip}$. 
In Fig. \ref{fig:4} we show the posterior distributions of the mass-radius relation of quark stars. 
We find that adding a crust results in a negligible effect on the quark star maximum mass but yields a little increase on the star radius (the increase depending on the chosen density $\varepsilon_{\rm crust}$). 
Quantitatively, the relative difference of radius between the quark stars with and without a crust is less than $1.9\%$ at $M = 1.4\Msun$.
We thus consider the crustal effects on the mass-radius relations of quark stars are indeed small and can be safely neglected in the present analysis. Nevertheless, it can be easily added in future works in the light of more accurate measurements.

\begin{table}
  \centering
  \caption{Most probable intervals of the EOS parameters ($90\%$ confidence level) in three families of EOSs (in the representative CFLm model with three $\varepsilon_{\rm crust}$ values) constrained by the joint MSP J0740+6620 and PSR J0030+0451 analysis (see details in Sec.~\ref{sect:obs}).}
\setlength{\tabcolsep}{8.pt}
\renewcommand\arraystretch{1.3}
  \begin{tabular}{lccc}    \hline \hline
    \multirow{2}*{Parameters}
   & \multicolumn{3}{c}{$\varepsilon_{\rm crust}/{\rm MeV\cdot fm^{-3}}$} \\
   \cline{2-4}
    {}& 0.24 & $10^{-2}$ &  $10^{-4}$     \\
    \hline 
    $B_{\rm eff}^{1/4}/{\rm MeV}$& $136.5_{-9.0}^{+10.3}$ &  $135.5_{-8.7}^{+10.6}$ &  $134.7_{-7.7}^{+10.8}$  \\
    \hline
    $a_4$ & $0.56_{-0.13}^{+0.20}$    
    & $0.54_{-0.12}^{+0.18}$ & $0.53_{-0.12}^{+0.20}$  \\
    \hline
    $\Delta/{\rm MeV}$ & $35.4_{-32.0}^{+44.3}$ &  $35.0_{-32.2}^{+45.4}$ & $34.4_{-30.7}^{+45.1}$ \\
    \hline
    $m_s/{\rm MeV}$ & $50_{-44}^{+73}$ &  $52_{-46}^{+66}$ 
    & $52_{-46}^{+69}$  \\
    \hline \hline
\end{tabular}
  \label{tb:3}
   \vspace{-0.2cm}
\end{table}

\vspace{0.5cm}
\section{Summary}

In conclusion, applying the Bodmer-Witten hypothesis, we here provide the first Bayesian analysis on the simultaneous high-accuracy measurements of mass and radius from {\it NICER} based on three types of quark star EOS parametrization, and study open problems of the maximum mass, the typical radius, etc.
We discuss in the context that all compact stars should be quark stars instead of neutron stars.
It is different from the two-families scenario in which neutron stars and quark stars can coexist.
It is also different from the scenario that the compact object is a hybrid star (neutron stars with quark matter in their interior)~\citep{2020Univ....6...81B,2020ApJ...904..103M,2021arXiv210315119L}, where the Bodmer-Witten hypothesis is not applied.

We treat the stability arguments of quark matter from standard nuclear physics as prior knowledge of the EOS ahead of an application of quark stars.
The mass distribution measured for the $2.14 \,\Msun$ pulsar, MSP J0740+6620, is used as the lower limit on the maximum mass.
We then provide the posterior probability distributions over the EOS model parameters and the quark star properties.
The {\it NICER} data of PSR J0030+0451 is found to support EOSs with an enhanced stiffness above that required to support the presently heaviest $2.14 \,\Msun$ pulsar.
The quark star maximum mass $M_{\rm TOV}$ is found to be in the range of $2.15-2.64 \Msun$, to the $90\%$ credibility interval. $R_{\rm 1.4}$ is centred around $12.3$ km, which is relatively smaller than the neutron star typical radius applying the same data.

A major caveat of our present analysis of the X-ray emission~\citep[see also in, e.g.,][]{2020ApJ...905....9T} assumes that quark stars have a similar crust as neutron stars. Although their effects on the mass and radius are found to be negligible, they may actually be bare or have a different crust regarding, e.g., the thickness, the composition, or even an atmosphere, which would change profoundly the X-ray modelling.
An improved study will require a consistent analysis of the {\it NICER} observations from the beginning with quark matter assumptions, which might be a considerable effort in the future.

\section*{ACKNOWLEDGMENTS}
The work is supported by National SKA Program of China (No.~2020SKA0120300), the National Natural Science Foundation of China (Grant No.~11873040), the science research grants from the China Manned Space Project (No. CMS-CSST-2021-B11) and the Youth Innovation Fund of Xiamen (No. 3502Z20206061).
\section*{Data Availability}

The data underlying this article are available in the article.

\end{document}